\documentclass[pra,aps,twocolumn,10pt,showpacs,groupedaddress,superscriptaddress,floatfix,notitlepage]{revtex4-2}
\usepackage{amsmath,amsfonts,amssymb,graphics,graphicx,epsfig,color,times}
\usepackage[utf8x]{inputenc}
\usepackage{color}
\usepackage{bbm, dsfont} 
\usepackage{subfigure}
\usepackage{hyperref}
\usepackage{mathrsfs}
\usepackage{verbatim}

\begin{document}
\title{Nonorthogonal coding in spectrally-entangled photons}
\author{N.-Y Tsai}
\affiliation{Department of Physics, National Taiwan University, Taipei 10617, Taiwan}

\author{H. H. Jen}
\email{sappyjen@gmail.com}
\affiliation{Institute of Atomic and Molecular Sciences, Academia Sinica, Taipei 10617, Taiwan}
\affiliation{Physics Division, National Center for Theoretical Sciences, Taipei 10617, Taiwan}

\date{\today}
\renewcommand{\r}{\mathbf{r}}
\newcommand{\f}{\mathbf{f}}
\renewcommand{\k}{\mathbf{k}}
\def\p{\mathbf{p}}
\def\q{\mathbf{q}}
\def\bea{\begin{eqnarray}}
\def\eea{\end{eqnarray}}
\def\ba{\begin{array}}
\def\ea{\end{array}}
\def\bdm{\begin{displaymath}}
\def\edm{\end{displaymath}}
\def\red{\color{red}}
\pacs{}
\begin{abstract}
Controlling and engineering continuous spectral modes of entangled photons represents one of the promising approaches toward secure quantum communications. By using the telecom bandwidth generated from a cascade-emitted biphoton in atomic ensembles, a fiber-based long-distance quantum communication can be feasible owing to its low transmission loss. With multiplexed photon pairs, we propose to implement a nonorthogonal coding scheme in their spectral modes and present an architecture of multiple channels enabling a high-capacity transfer of codewords. Using the measures of the second-order correlations and associated visibility and contrast, we further quantify the performance of the proposed nonorthogonal coding scheme. Our results demonstrate the capability to encode and decode quantum information beyond the orthogonal coding scheme, and in particular, the multi-channel setup manifests a resilience and an advantage in a design with multiple channel errors. The proposed scheme here can be applicable to a large-scale and multiuser quantum communication and pave the way toward an efficient and functional quantum information processing. 
\end{abstract}
\maketitle

\section{Introduction} 

A secure quantum communication can be ensured by use of quantum cryptography \cite{Gisin2002} and quantum key distributions \cite{Scarani2009, Xu2020}. This complete security relies on the fundamental characteristics of quantum mechanics, where the uncertainty principle posits a limit on the accurate determination of conjugate observables simultaneously and the communication system cannot be found undisturbed under a measurement. The former allows two parties to share a random secret key, while the latter enables them to detect any eavesdropping from another party. To facilitate such secure communication, photons are great candidates to carry quantum information, while atoms are best serving as local storage of it. An efficient quantum interface between light and matter \cite{Hammerer2010} thus provides the foundation to relay quantum information as in quantum repeaters \cite{Dur1999, Duan2001}, which further promises a large-scale quantum network \cite{Cirac1997, Kimble2008} and scalable quantum computation.  

One essential aspect of scalability in quantum science and technology using flying photonic qubits, in addition to the number of them, regards the multiple degrees of freedom that can be accessed in photons. These degrees of freedom for encoding and decoding quantum information can be discrete as light polarizations \cite{Clauser1969, Aspect1981, Kwiat1995, Crespi2011, Wang2019} and frequency-bin qubits \cite{Lu2018, Kues2019, Mahmudlu2023} or continuous as transverse momenta \cite{Law2004, Moreau2014}, space \cite{Grad2012}, orbital angular momenta \cite{Arnaut2000, Mair2001, Molina2007, Dada2011, Fickler2012, Nicolas2014, Ding2015}, and spectral modes of light \cite{Branning1999, Law2000, Parker2000, Lukens2014, Jen2016, Lukens2017, Yang2021}. With continuous-variable entanglement of optical modes in a multiplexing architecture \cite{Pu2017, Shi2020}, a potential high-capacity quantum resource can be feasible, which gives rise to an enhanced performance in multimode quantum communication and high-dimensional quantum information processing \cite{Braunstein2005}. Moreover, the spectral encoding is robust and stable against decoherence, which can be incorporated with integrated photonic elements to be compact and scalable in the on-chip photonic platforms \cite{Yang2021, Mahmudlu2023}.  

In spite of using continuous modes of light as high-capacity quantum channels in quantum communication, an efficient light transmission is as well crucial and demanded, for example, in a fiber-based communication network \cite{Liao2017, Liao2018, Chen2021}. This can be achieved by utilizing the telecommunication bandwidth of light to minimize its transmission loss, which can be acquired in a cascade-emitted biphoton generated from an atomic ensemble \cite{Chaneliere2006, Radnaev2010, Jen2010}, serving as an excellent frequency-entangled state between the telecom and infrared bandwidths. The telecom photon suffices long-distance quantum communication, while the infrared can be stored locally as a collective spin wave \cite{Chaneliere2005, Chen2006} with high efficiency \cite{Chen2013, Yang2016, Hsiao2018}. In this highly-correlated biphoton source, an almost pure single photons can be heralded by removing continuous frequency entanglement among them \cite{Seidler2020, Wong2021, Wong2022}, which is useful and can be applied in quantum computation with linear optics \cite{Knill2001}, optical quantum network implementation \cite{Dusanowski2019}, and realizing photon${\rm -}$photon quantum logic gates \cite{Li2021}.  

Here we propose to implement a nonorthogonal coding scheme in spectrally-entangled photons, where the biphoton source can be generated from an atomic diamond configuration under a four-wave mixing process \cite{Radnaev2010, Jen2017, Wong2022}. The orthogonal coding of Hadamard codes has been implemented spectrally in the broadband entangled photons through spontaneous parametric down conversion, where a matched code recovers a sharp and narrow correlation peak \cite{Lukens2014}. To further go beyond the orthogonal coding scheme, we utilize the quasi-orthogonal space–time block code \cite{Alamouti1998, Tarokh1999, Jafarkhani2001}, a well developed technique in wireless classical communication. This block code is originally designed to transmit multiple and redundant copies of data stream across many antennas in a reliable manner even under an influence of channel transmission issues or data corruptions. Under the conditions of high transmission rates with low signal-to-noise ratios, the quasi-orthogonal code can outperform the orthogonal one in reducing the transmission bit-error rates \cite{Jafarkhani2001}. 

The meaning of ‘nonorthogonal’ in the coding space represents the property that the codes are not orthogonal to each other. One of the orthogonal codes is the Hadamard code which has been demonstrated in the frequency bins \cite{Lukens2014}. This is a natural choice since the code spaces are distinguishable to each other and show enhanced correlations when they are matched in the decoding stage. On the other hand, a general coding space can be nonorthogonal, which essentially generalizes the coding process. In this work, we propose a nonorthogonal coding scheme which can be facilitated in a multiplexing platform of biphotons from multiple atomic ensembles, where the information of codewords is encoded in the spectral modes determined by Schmidt decompositions. We further propose a design of multiple $R$ channels within each $M$ photon pairs are multiplexed to encode the nonorthogonal codes. We quantify its performance by calculating the second-order correlation functions along with useful measures of visibility and contrast, up to a capacity of coding space dimension of $M^R$. We find that there can be a balance between the contrasts in the correlation functions and the code space dimensions, where a potentially best design can be identified in our multi-channel nonorthogonal coding scheme. Lastly, we present the advantages of the nonorthogonal coding scheme in multiple channels, which is shown to be more resilient to channel errors than the conventional orthogonal coding scheme which can only be implemented in a single-channel setup. Our results can shed new light on multiuser long-distance quantum communication with high capacity of coding channels and provide insights to a blueprint of scalable quantum network.  

\section{Correlation function of entangled photon pairs}

\begin{figure}[t]
\includegraphics[width=0.48\textwidth]{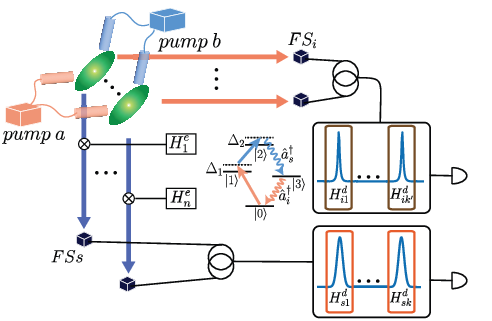}
\caption{A multiplexed coding scheme. A number of $n$ atomic ensembles are illustrated to generate $n$ photon pairs, where an inset plot shows the four-level atomic configuration with a single- and two-photon detuning, $\Delta_1$ and $\Delta_2$, respectively. The pump fields drive the atomic transitions $|0\rangle\rightarrow|1\rangle$ and $|1\rangle\rightarrow|2\rangle$, respectively, which lead to subsequently and spontaneously emitted signal photon $\hat a_s$ and idler photon $\hat a_i$. The encoding operations $H^e_1$ to $H^e_n$ act on individual signal photons, which are frequency-multiplexed together with the corresponding idler photons by frequency shifters (FS), followed by the decoding operations $H^d_{s1}$ to $H^d_{sk}$ and $H^d_{i1}$ to $H^d_{ik'}$ on the signal and idler photons, respectively. A single-channel scheme represents the case when $k=k'=n$, while in general a multi-channel scheme allows $k\neq k'$ and $k,k'<n$, leading to a huge code space in the nonorthogonal coding scheme.}
\label{fig1:multiplexCoding} 
\end{figure}

We first introduce the spectral function of the entangled photon pairs generated via the four-wave mixing process. These photon pairs can be generated in a cold atomic ensemble with a four-level structure as shown in Fig. \ref{fig1:multiplexCoding}. Two classical excitation pulses drive the atoms and lead to the cascaded emissions of signal and idler photons, which are highly correlated in their propagating directions. The reason why the cascade atomic transition considered in this work is for the potential application in fiber-based quantum communication, where the low-loss telecom wavelengths can be generated in its upper transition ($6S$ or $4D$ to $5P$ transitions for rubidium atoms as examples), while its lower transition ($5S$ to $5P$ transition) is in the infrared band suitable for local storage in quantum memory. This biphoton source fit both needs of the low-loss quantum information transmission in a fiber-based quantum network and the local qubit storage for entanglement distribution \cite{Chaneliere2006, Radnaev2010}. Meanwhile, the biphoton source generated from nonlinear crystals can also be suitable for the purpose of encoding and decoding of quantum information, but the photons usually reside in the same central frequency. 

The effective spectrally-entangled biphoton state can be expressed in continuous frequency spaces as \cite{Jen2016, Jen2017} 
\begin{equation}
    |\Psi_b\rangle = \frac{1}{\mathcal{N}} \int \int f(\omega_s, \omega_i)\hat{a}_s^{\dagger} \hat{a}_i^{\dagger}|0,0\rangle d\omega_s d\omega_i,
\end{equation}
with $\hat a_{s(i)}^\dagger$ representing the signal (idler) photon creation operator, a normalization factor $\mathcal{N}$, and the biphoton spectral function as 
\begin{equation}
    f(\omega_s, \omega_i) = \frac{e^{-(\Delta \omega_s+\Delta \omega_i)\tau^2/8}}{\Gamma_3^N/2-i\Delta \omega_i}, \label{f}
\end{equation}
where the respective detunings for the signal and the idler photons are \(\Delta \omega_s \equiv \omega_s - \omega_2 +\omega_3 -\Delta_2\) and \(\Delta \omega_i \equiv \omega_i - \omega_3\), and the detail of the above derivation can be found in the Appendix. The spectral function represents the probability amplitude and is proportional to the generation rate of this biphoton source. Under the assumption of laser pulses in Gaussian forms, the joint Gaussian distribution in Eq. (\ref{f}) conserves the signal and idler photon central energies as $\omega_s+\omega_i=\omega_a+\omega_b$ with a spectral width approximately determined by $1/\tau$, the inverse of laser pulse duration, which indicates an entangling origin in contrast to a separate idler distribution in a Lorentzian profile with a width of superradiant decay constant $\Gamma_3^N$.  

The continuous entanglement entropy in the biphoton state can be analyzed through Schmidt decompositions \cite{Law2000, Parker2000}, where the state can be reinterpreted as \(|\Psi_b\rangle=\sum_{n=1}^\infty \sqrt{\lambda_n}b_n^{\dagger} c_n^{\dagger} |0,0\rangle\) with a probability $\lambda_n$ under the orthogonal Schmidt modes $\psi_n(\omega_s)$ and $\phi_n(\omega_i)$, that is \(b_n^{\dagger}=\int \psi_n(\omega_s) \hat a_s^\dag(\omega_s)d\omega_s\) and \(c_n^{\dagger}=\int \phi_n(\omega_i)\hat a_i^\dag(\omega_i) d\omega_i\), respectively. A normalization of $\sum_{n=1}^\infty\lambda_n=1$ should be satisfied and the entanglement entropy can be calculated as $S=-\sum_{n=1}^\infty\lambda_n\ln\lambda_n$, which leads to an entangled bipartite source as long as $\lambda_1\neq 1$, otherwise it suggests a separable state with $S=0$. 

Next we obtain the second-order correlation function, which is proportional to the coincidence rate and is essential in determining the outcome of the performance in the encoding and decoding scheme we propose here. It can be calculated as \cite{Scully1997}
\bea
    g^{(2)}(t'-t) =&& \frac{\langle \hat{E}_s^{(-)}(t) \hat{E}_i^{(-)}(t') \hat{E}_i^{(+)}(t') \hat{E}_s^{(+)}(t) \rangle}{\langle \hat{E}_s^{(-)}(t)\hat{E}_s^{(+)}(t) \rangle\langle \hat{E}_i^{(-)}(t) \hat{E}_i^{(+)}(t) \rangle}, \\
    \hat{E}_s^{(+)}(t) =&& \sum_{k_s, \lambda_s} \sqrt{\frac{\hbar \omega_s}{2\epsilon_0 V}} \hat{a}_{k_s, \lambda_s} \vec{\epsilon}_{k_s, \lambda_s} e^{-i \omega_s t}, \\
    \hat{E}_i^{(+)}(t) =&& \sum_{k_i, \lambda_i} \sqrt{\frac{\hbar \omega_i}{2\epsilon_0 V}} \hat{a}_{k_i, \lambda_i} \vec{\epsilon}_{k_i, \lambda_i} e^{-i \omega_i t},
\eea
where $\hat E_{s}^{(+)}$ and $\hat E_{i}^{(+)}$ represent the vector forms of the signal and the idler electric fields, respectively, with $k_{s}$, $k_i$ denoting the wave vectors and $\lambda_s$, $\lambda_i$ the field polarizations. We assume that the photon pairs are transmitted though single-mode fibers. Therefore the spatial correlations can be neglected and we focus only on their frequency correlations. Here we consider \(t'=t\rightarrow \infty\) and then obtain \(g^{(2)}(0)\) under the Schmidt bases as
\begin{equation} \label{eqn:g2}
    \begin{split}
    g^{(2)}(0) =& \int \Bigg|\sum_{n=1}^\infty \sqrt{\lambda_n} (\psi_n*\phi_n)(\omega)\Bigg|^2 d\omega, \\
    (\psi_n*\phi_n)(\omega) =& \int \psi_n(\omega_s)\phi_n(\omega-\omega_s) d\omega_s,
    \end{split}
\end{equation}
where the zero-time second-order correlation function is equivalent to the integral of convolutions between the signal and idler photon eigenmodes in frequency spaces.

\section{Main results}

Here we show our main results of the second-order correlations $g^{(2)}(0)$ from the proposed scheme as shown in Fig. \ref{fig1:multiplexCoding}, where we can quantify its performance by the contrast of $g^{(2)}(0)$. For both a single-channel and multi-channel nonorthogonal coding scheme, we multiplex the signal and idler photon pairs by frequency shifters, leading to well-separated spectral modes which are feasible for individual mode addressing. The multiplexed spectral function by frequency shifters can be expressed in general as
\bea
    f_M(\omega_s,\omega_i) =&&\sum_{n=1}^{N_{\rm ph}} H_n \frac{e^{-(\Delta\omega_s + \Delta\omega_i+\delta_{qn})^2 \tau^2 /8}}{\Gamma_3^N /2 - i(\Delta\omega_i-\delta_{pn})},\\
		\equiv&& \sum_{n=1}^{N_{\rm ph}} H_nf_n(\omega_s,\omega_i),
\eea
where $H_n$ represents the multiplexed weights on the individual photon pairs with \(\delta_{pn}\) the frequency shift in the idler photon and \(\delta_{qn}\) the joint frequency shift in the signal and idler photons. $H_n$ can be controlled by external laser driving parameters as in Eq. (\ref{biphoton}), which can be treated further as the encoding operations $H_n=H_n^e$ in Fig. \ref{fig1:multiplexCoding}. Throughout the paper, we arrange a number of $N_{\rm ph}$ multiplexed photon pairs in parallel with an energy-conserved axis satisfying \(\Delta \omega_s +\Delta \omega_i = 0\), corresponding to anti-correlated signal and idler photons without loss of generality. 

Under this frequency-separated multiplexed scheme, the first $N_{\rm ph}$ eigenmodes determined by Schmidt decompositions in a descending order of $\lambda_n$ become almost degenerate and dominate over the other modes. We then can approximate the eigenmodes in Eq. (\ref{eqn:g2}) by using the integrated profiles of signal and idler photons, respectively, 
\bea
    \psi_n(\omega_s) \propto&& \int_{-\infty}^\infty f_n(\omega_s,\omega_i) d\omega_i, \nonumber\\
		=&& -\frac{1}{N_s} e^{-(\Delta \omega_s + \delta_{pn} +\delta_{qn} +i\Gamma_3^N/2)^2 \tau^2 /8},\\
    \phi_n(\omega_i) \propto&& \int_{-\infty}^\infty f_n(\omega_s,\omega_i) d\omega_s,\nonumber\\
		=&&\frac{1}{N_i}\frac{1}{\Gamma_3^N/2 - i(\Delta \omega_i-\delta_{pn})},
\eea
where \(N_s, N_i\) are normalization factors for these $N_{\rm ph}$ signal and idler eigenmodes with frequency shifts $-(\delta_{pn} +\delta_{qn})$ and $\delta_{pn}$ in Gaussian and Lorentzian profiles, respectively. The convolution in Eq. (\ref{eqn:g2}) can further be obtained as 
\begin{equation} \label{eqn:g2convolutionreduced}
    (\psi_n*\phi_n)(\omega) = \frac{1}{N_s N_i} e^{-(\omega+\delta_{qn})^2\tau^2/8},
\end{equation}
which depends only on the joint frequency shift \(\delta_{qn}\). This leads to \(g^{(2)}(0)=2\sqrt{\pi}/(N_s^2N_i^2\tau)\) in a well-separated multiplexed scheme as $\delta_{qn}\gg \delta_{qn'}$. Next we introduce the nonorthogonal coding under this multiplexed scheme. 

\subsection{Nonorthogonal multiplexed coding scheme}

First, we consider the multiplexed photon pairs under \(\delta_{qn}=0\), where spectrally-entangled  signal and idler photons locate along the energy-conservation axis and we denote it as a single channel coding scheme. In this way, the encoding operations are denoted as \(H_m^e\) on the signal photons, and the decoding ones on the idler photons are represented by \(H_{im}^d\) for $m=[1,N_{\rm ph}]$ with \(H^d_{sm}=1\) in Fig. \ref{fig1:multiplexCoding}. The coding operations in general involve both the phase and amplitude modulations on the respective photon pairs, which can be facilitated by a phase imprinting technique and beam splitters. We can approximate \(g^{(2)}(0)\) by considering only the first $N_{\rm ph}$ modes as in Eq. (\ref{eqn:g2convolutionreduced}), and from Eq. (\ref{eqn:g2}) we obtain  
\begin{equation} \label{eqn:g2reduced}
    g^{(2)}(0) \approx\frac{2\sqrt{\pi}}{N_s^2N_i^2\tau N_{\rm ph}} \left|\sum_{n=1}^{N_{\rm ph}} H_{in}^d H_n^e\right|^2,
\end{equation}
where the Schmidt mode numbers or the mode probability $\lambda_n\approx 1/N_{\rm ph}$. When the encoding and decoding codes are matched, \(g^{(2)}(0)\) manifests a maximum, while it drops significantly in the mismatched case, for example in the orthogonal Hadamard codes \cite{Lukens2014}. 

\begin{figure}[t]
\centering
\includegraphics[width=0.48\textwidth]{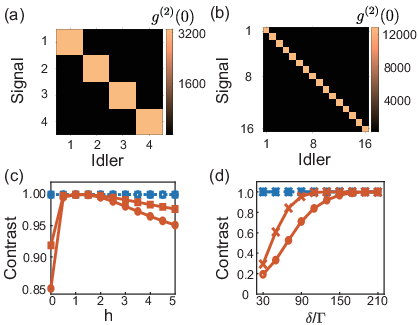}
\caption{\label{fig2:singlechannel} Second-order correlation $g^{(2)}(0)$ in a single-channel nonorthogonal coding scheme. The ideal spectral decoding performance of $g^{(2)}(0)$ for (a) $N_{\rm ph}=4$ and (b) $16$, respectively, where $\delta_{pn}\rightarrow \infty$ and $\lambda_{n\in[1,N_{\rm ph}]}=1/\sqrt{N_{\rm ph}}$. The parameters of $\vec{c}$ is chosen by setting $h=2$. As a comparison in the ideal case, $V$ (blue-$\square$ and -$\circ$) and $C_{\rm od}$ (red-$\square$ and -$\circ$) are plotted as $h$ varies in (c), where $N=4$ ($\square$) and $16$ ($\circ$), respectively. In (d), $V$ (blue-$\circ$ and -$\times$) and $C_{\rm od}$ (red-$\circ$ and -$\times$) are plotted as $\delta$ varies for $N = 4$ and $h=1$, with the frequency shifts of photon pairs \(\delta_{p1} = -1.5\delta, \delta_{p2} = -0.5\delta, \delta_{p3} = 0.5\delta, \delta_{p4} = 1.5\delta\). The excitation pulse durations are chosen as \(\Gamma\tau = 0.25\) ($\circ$) and \(0.4\) ($\times$), respectively, for comparisons. We consider \(\Gamma/2\pi = 6\)MHz as an example for D$_1$ transition of rubidium atoms and $\Gamma_3^N/\Gamma=5$.}
\end{figure}

To go beyond the orthogonal coding scheme, here we introduce the quasi-orthogonal space-time block codes, which have been useful in multiuser antenna systems. This nonorthogonal code can be constructed from the so-called Alamouti \(2\times2\) codes~\cite{Alamouti1998,Tarokh1999,Jafarkhani2001}, 
\begin{equation}
    C_{1,2} = 
    \begin{pmatrix}
        c_1 & c_2\\
        -c_2^{*} & c_1^{*} 
    \end{pmatrix},
\end{equation}
from which Alamouti \(N_a \times N_a \) codes can be generated by an extension of Alamouti form with arbitrary $c_1$ and $c_2$. Therefore, the nonorthogonal codes we apply here are not unique. We will investigate the non-uniqueness of the nonorthogonal codes in Fig. \ref{fig2:singlechannel} in the following. In terms of the contrast in the photon-photon correlations, the performance of using the nonorthogonal codes does not exceed the one with the orthogonal codes. Meanwhile, the advantage of the nonorthogonal codes lies at the design of multi-channel scheme where the photon correlations are robust to channel errors, as will be shown in Fig. \ref{Fig5}.

As a demonstration, we take $N_a = 4$ as an example, which reads
\begin{equation}
    C_{1,4} = 
    \begin{pmatrix}
        C_{1,2} & C_{3,4}\\
        -C_{3,4}^{*} & C_{1,2}^{*} 
    \end{pmatrix}
    =
    \begin{pmatrix}
        c_{1} & c_{2} & c_3 & c_4\\
        -c_{2}^{*} & c_{1}^{*} & -c_4^* & c_3^* \\
        -c_3^* & -c_4^* & c_{1}^* & c_{2}^* \\
        c_4 & -c_3 & -c_{2} & c_{1} 
    \end{pmatrix}. \label{C14}
\end{equation}
In the above, the orthogonal relations only hold between the columns $(1,2)$, $(1,3)$, $(2,4)$, and $(3,4)$, and again $c_3$ and $c_4$ are arbitrary. In general, the nonorthogonal codes can be constructed as 
\begin{equation}
    C_{1,N} = 
    \begin{pmatrix}
        C_{1,N/2} & C_{N/2+1,N}\\
        -C_{N/2+1,N}^{*} & C_{1,N/2}^{*} 
    \end{pmatrix},\label{C}
\end{equation}
with an even $N$, where $C_{N/2+1,N}$ denotes a similar construction of $C_{1,N/2}$ with $c_j\leftrightarrow c_{N/2+j}$ for $j=[1,N/2]$.  

We can then implement the encoding \(H^e_m\) and decoding operations \(H^d_m\) using the column vectors of Eq. (\ref{C}) as the nonorthogonal coding scheme. We note that the choice of \(\vec{c} = (c_1,c_2,..., c_N)\) from \(c_l = ar^{l-1}\) with arbitrary \(a\) and \(r\), for example, can lead to an orthogonal coding scheme. Therefore, in general the nonorthogonal coding scheme involves an orthogonal one as a special case. To quantify the coding performance in terms of contrasts in correlations, we define the visibility \(V=(g^{(2)}_{\rm max}-g^{(2)}_{\rm min})/(g^{(2)}_{\rm max}+g^{(2)}_{\rm min})\) and the off-diagonal contrast \(C_{\rm od}=(g^{(2)}_{\rm max}-g^{(2)}_{\rm od})/(g^{(2)}_{\rm max}+g^{(2)}_{\rm od})\), where we neglect the zero-time dependence for concise expressions. \(g^{(2)}_{\rm max}\) and \(g^{(2)}_{\rm min}\) are global maximum and minimum in $g^{(2)}(0)$, respectively, while \(g^{(2)}_{\rm od}\) represents the largest correlation element in the mismatched cases in the odd-diagonal (od) $g^{(2)}(0)$. This provides a way for distinguishing coding capability and its performance, under different choices of $\vec{c}$ and various system parameters which we can control and manipulate in our proposed multiplexing scheme.   

In Fig. \ref{fig2:singlechannel}, we present the second-order correlations in a single-channel nonorthogonal coding scheme. In Figs. \ref{fig2:singlechannel}(a) and \ref{fig2:singlechannel}(b), we implement the nonorthogonal coding scheme, where \(\vec{c}\) is chosen as to be $N_{\rm ph}$ equally-spaced points between $[1,h=2]$. In the ideal case when $\delta_{pn}$ is large enough that the encoding operates on the well-separated spectral modes, we utilize Eq. (\ref{eqn:g2reduced}) to demonstrate the ultimate high correlations when the codes are matched and their dependence on the number of entangled photon pairs in Figs. \ref{fig2:singlechannel}(a) and \ref{fig2:singlechannel}(b). To further demonstrate the effect of free parameters $\vec{c}$ in the nonorthogonal coding scheme, we show $V$ and $C_{\rm od}$ in Fig. \ref{fig2:singlechannel}(c) as $h$ varies. At $h=1$, \(\vec{c} =(1,1,...,1)\), which reduces the nonorthogonal codes to the orthogonal ones as Hadamard codes. $V$ stays maximized since $g^{(2)}_{\rm min}$ can be vanishing in the nonorthogonal coding scheme owing to its quasi-orthogonality property. On the other hand, $C_{\rm od}$ is maximized for the Hadamard codes, whereas the nonorthogonal codes can still maintain significant contrasts within our considered range of $h>1$. Below $h=1$ in Fig. \ref{fig2:singlechannel}(c), $\vec{c}$ is chosen within $[h,1]$ and is more prone to the nonorthogonal nature than in the range of $h>1$. Therefore, $C_{\rm od}$ suffers from and decreases more for a larger $N_{\rm ph}$. We note that the overall scaling of $\vec{c}$ does not matter in the contrasts, and therefore $C_{\rm od}$ would be the same according to the condition of $hh'=1$ for $\vec{c}$ chosen between $[h'<1,1]$ and $[1,h>1]$, respectively. This can be seen and expected in Fig. \ref{fig2:singlechannel}(c).    

In Fig. \ref{fig2:singlechannel}(d), we further investigate the impact from the spectral interferences between each multiplexed spectral modes. We take $N_{\rm ph}=4$ as an example with a frequency coding bin with a range of \(\delta\). As $\delta$ increases, the multiplexed modes approach the ideal case for $C_{\rm od}$ as in Fig. \ref{fig2:singlechannel}(c), where $C_{\rm od}\rightarrow 1$. When $\delta$ is finite, these multiplexed modes are subject to finite spectral interferences, leading to a suppression of $C_{\rm od}$. This can be attributed to imperfect encoding and decoding operations in frequency spaces, contrary to the assumption of well-separated frequency modes. As we vary the excitation pulse duration $\tau$, a shorter pulse manifests a broader spread of frequency correlations, which enhances the spectral interferences and gives rise to the suppression of $C_{\rm od}$. The visibility $V$ here stays maximized in our considered large range of $\delta$, which is less sensitive to the effect of spectral interferences than $C_{\rm od}$. As shown in Figs. \ref{fig2:singlechannel}(c) and \ref{fig2:singlechannel}(d), the nonorthogonal coding scheme can be reduced to the orthogonal one, exactly the same as the Hadamard codes when a range of $[1, h=1]$ is used for $N$ coding spaces, where the photon correlations approach the results of the ideal cases as in well-separated coding spaces in frequency bins as in the orthogonal codes \cite{Lukens2014}.

\subsection{Multiple channels}

\begin{figure}[t]
\centering{}
\includegraphics[width=0.48\textwidth]{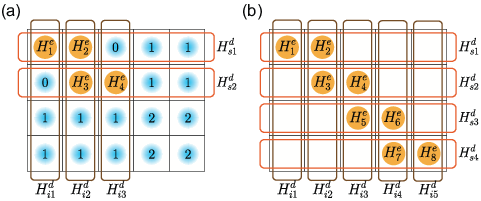}
\caption{\label{fig3:multiple_Design} The decoding operations in multiple channels. Yellow filled circles represent the photon pairs encoded by $H^e_n=H^e_{rm}$ where $n=r+(m-1)R$, under the multiplexed scheme in a frequency-separated domain. The decoding operations $H^{d}_{sk}$ and $H^{d}_{ik'}$ on the signal and idler photons, respectively, result in effectively the decoding operation $H^d_{rm}=H^{d}_{sk}H^{d}_{ik'}$ on the spectrally-entangled modes of the photon pairs. The numbers in blue circles in (a) indicate how many degrees of freedom left in $H^{d}_{sk}$ or $H^{d}_{ik'}$ that can be added for decoding operations. The photon pairs cannot be placed at the multiplexed domain with the number equal to $0$, giving the constraint for the multiplexed photon pairs in a multi-channel structure. In (b), an example of $M=4$ photon pairs each for $R=2$ channels shows one of the dense structures of code spaces with only one redundant degree of freedom in either $H^{d}_{sk}$ or $H^{d}_{ik'}$.}
\end{figure}

Here we extend the previous single-channel setup to a multi-channel platform by releasing the requirement of \(\delta_{qn}=0\). Multiple channels can then be constructed by well separating \(\delta_{qn} = \delta_r\) as shown in Eq. (\ref{eqn:g2convolutionreduced}), where each channel with specific $\delta_r$ does not crosstalk with each other when \(\delta_{r}\) is large enough with vanishing spectral interferences. Under this ideal condition, the correlation function for $R$ channels and $M$ multiplexed photon pairs in each channel in general can be obtained as 
\bea\label{eqn:g2multi}
    g^{(2)}(0)=&&\frac{1}{N_s^2N_i^2}\int \left|\sum_{r=1}^R \sum_{m=1}^M \sqrt{\lambda_{rm}} e^{-(\omega+\delta_{r})^2\tau^2/8}\right|^2 d\omega, \nonumber\\
    =&& \frac{2\sqrt{\pi}}{N_s^2N_i^2\tau M} \sum_{r=1}^R \left|\sum_{m=1}^M H_{rm}^e H_{rm}^d\right|^2, 
\eea
where the decoding operation $H_{rm}^d$ requires both signal and idler coding operations $H^d_{sk}$ and $H^d_{ik'}$ respectively, as shown in Fig. \ref{fig3:multiple_Design}. The encoding and decoding operations $H_{rm}^e$ and $H_{rm}^d$ correspond to $H_{n=(r-1)M+m}^e$ and \(H_{sk}^dH_{ik'}^d\), respectively. The total number of photon pairs are \( R M\), and each code subspace in the $r$th channel can be constructed by the nonorthogonal code with a size $M$. In this design, spectrally-entangled modes in different channels allow for overlapped spectral coding, with an advantage of saving the limited frequency spaces. When encoding and decoding operations are matched in respective channels, \(g^{(2)}(o)\) will reach its maximum. If $R'$ channels are considered, \(g^{(2)}(o)\) will be enhanced \(R'\) times as the matched case in a single-channel setup. In the multi-channel design in Fig. \ref{fig3:multiple_Design}, the code space dimension will become $M^{R}$. This dimension grows exponentially to reach a maximum when an even $M$ is chosen as $2$. 

There are some constraints on the frequency-multiplexed photon pairs in the design of Fig. \ref{fig3:multiple_Design}. To achieve the encoding on the entangled modes which overlap in the signal or idler spectra, the design should ensure that any decoding operations \(H^d_{rm}\) can successfully reflect the nonorthogonal coding sequences. Therefore, in Fig. \ref{fig3:multiple_Design}(a), we present a constraint on placing or multiplexing the photon pairs for encoding and decoding operations. The numbers in empty sites of potentially multiplexed photon pairs show the degrees of freedom that can be allowed by adding extra operations of $H^d_{sk}$ and $H^d_{ik'}$. The place with a null number presents a forbidden scenario in the design that the nonorthogonal codes we utilize cannot be achieved. One of the dense structures in frequency spaces for photon pairs is shown in Fig. \ref{fig3:multiple_Design}(b) for an even $M$, where we denote it as the staircase designing structure.

\begin{figure}[t]
\centering{}
\includegraphics[width=0.48\textwidth]{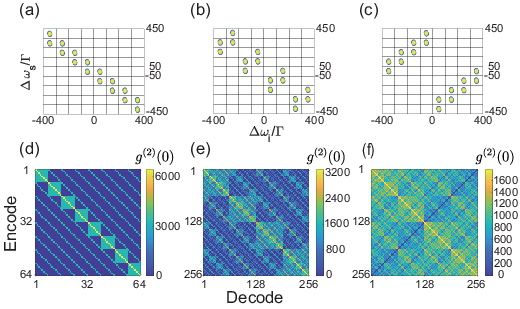}
\caption{\label{fig4:Multiple_channel} Multi-channel structure and $g^{(2)}(0)$. Three possible dense designing arrangements of photon pairs for $N_{\rm ph}=16$ are shown in (a-c), where $R=2$, $4$, and $8$ with $M = 8$, $4$, and $2$, respectively. We plot schematically by using $\delta = 100\Gamma$ to avoid significant $C_{non}$ caused by spectral interferences from adjacent photon pairs. Corresponding correlations $g^{(2)}(0)$ are calculated in (d), (e), and (f), respectively, under the ideal spectral decoding scheme. The code space dimensions are $64$, $256$, and $256$, respectively. The $\Gamma\tau$ is chosen as $0.5$.}
\end{figure}

In Fig. \ref{fig4:Multiple_channel}, we demonstrate the multi-channel platform with $16$ photon pairs in a multiplexing scheme in three different arrangements, which are the dense forms within the same frequency domains. We illustrate $2$, $4$, and $8$ channels with $8$, $4$, and $2$ photon pairs in each channel, respectively. Each channel encodes the nonorthogonal code subspaces \(C_{1M}\) with \(\vec{c} = (1, M/(M-1),...,2)\). \(\delta_r\) are chosen as $100$\(\Gamma\) to avoid the spectral interference as shown in Eq. (\ref{eqn:g2multi}) with the FWHM \(\approx 9\Gamma\) under a short pulse duration \(\Gamma\tau = 0.5\). 

Figures \ref{fig4:Multiple_channel}(a), \ref{fig4:Multiple_channel}(b), and \ref{fig4:Multiple_channel}(c) with $\delta=100\Gamma$ ensure that the spectral coding is well separated. This leads to the ideal coding operations in Figs. \ref{fig4:Multiple_channel}(d), \ref{fig4:Multiple_channel}(e), and \ref{fig4:Multiple_channel}(f), where we obtain $g^{(2)}(0)$ for comparisons. The same color legend indicates the same number of matched channel for our nonorthogonal codes. For example, in Fig. \ref{fig4:Multiple_channel}(d), $g^{(2)}(0)$ involves three kinds of finite values, which respectively correspond to $2$, $1$, and $0$ channels matched cases. As a result, an increase of number of channel will lead to $g^{(2)}(0)$ with more distinctive finite values. A comparison between Figs. \ref{fig4:Multiple_channel}(e) and \ref{fig4:Multiple_channel}(f) shows that although they can both achieve the capacity of codes to a dimension of $256$, the arrangement of $4$ channels case has less number of finite values involved in $g^{(2)}(0)$ than that of $8$ channels case. For more channels involved, the contrast $C_{non}=1/(2R-1)$ becomes less significant. This can be seen in Fig. \ref{fig4:Multiple_channel}(f), where a reduced contrast emerges compared to the cases in Figs. \ref{fig4:Multiple_channel}(d) and \ref{fig4:Multiple_channel}(e). Therefore, it would be optimal to have minimal possible channels in a multi-channel scheme with a greater contrast, as well as requiring the largest code space dimension that can be achieved for high-capacity transmission. As a demonstration, Fig. \ref{fig4:Multiple_channel}(b) stands out as the best design in the balance between the contrasts in $g^{(2)}(0)$ and the code space dimensions. In the results of $g^{(2)}(0)$, they can reach approximately the maximum of $6400$, $3200$, and $1600$ under the matched cases in Figs. \ref{fig4:Multiple_channel}(d), \ref{fig4:Multiple_channel}(e), and \ref{fig4:Multiple_channel}(f), respectively, which can be explained by Eq.(\ref{eqn:g2multi}) associated with the normalization factor $M$. For a fixed number of photon pairs in each channel, a code space capacity can grow exponentially as $M^R$ with an increasing channel number $R$, but nonetheless with a price of a reduced contrast in overall $g^{(2)}(0)$.

\subsection{Advantages of multi-channel nonorthogonal coding scheme}

\begin{figure}[t]
\centering{}
\includegraphics[width=0.48\textwidth]{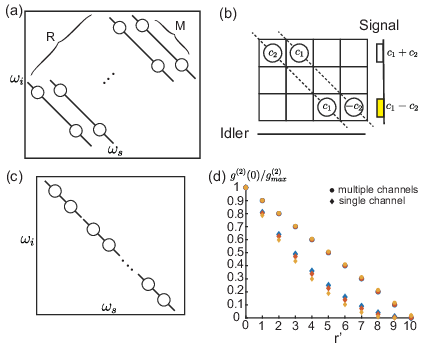}
\caption{\label{Fig5} Advantages of multi-channel nonorthogonal coding. (a) A schematic illustration for multi-channel setup with $N_{\rm ph}=MR$ photon pairs in total and $M=2$ in each of the $R$ channels, which forms a total code space of $M^{R}$. This multi-channel setup cannot be realized in a conventional orthogonal coding scheme owing to the overlapped spectral regions in the multiplexing scheme as shown in (b), where an unwanted cancellation shows up when $c_1=c_2=1$ using Hadamard codes in the signal photons. As a comparison, we design a corresponding single-channel setup with the same total code space in (a) but with $R$ segments instead as shown in (c). A relative reduction of $g^{(2)}(0)$ is compared in (d) for $N_{\rm ph}=40$ and $M=4$ versus various $r'$ channels or segments with errors. For the single-channel setup, three cases are compared with orthogonal and mismatched $r'$ segments with $h=2$ (blue-$\blacklozenge$), nonorthogonal and mismatched $r'$ segments with $h=2$ (red-$\blacklozenge$) and $h=4$ (yellow-$\blacklozenge$), respectively, as $g^{(2)}(0)$ descends at the same $r'$. By contrast in the multi-channel setup, the reductions in $g^{(2)}(0)$ are less severe and they are almost on top with each other in the corresponding mismatched cases ($\bullet$).}
\end{figure}

Here we discuss more on the advantages of multi-channel nonorthogonal coding scheme, which are associated with less contrasted $g^{(2)}(0)$ we have presented in the previous subsection. As we have emphasized that nonorthogonal coding scheme is preferential for multiuser communications, we explicitly demonstrate in Fig. \ref{Fig5} that the nonorthogonal coding scheme, especially in multiple channels, is robust to channel errors in addition to the allowed code space capacity. 

First of all, we note that the multi-channel design in Fig. \ref{Fig5}(a) with nonorthogonal coding utilizes spectrally-entangled photon pairs which are multiplexed with economical frequency ranges. As a comparison, within the same frequency ranges for encoding and decoding, the proposed multi-channel nonorthogonal coding scheme allows more code spaces than the conventional orthogonal coding scheme using divided and energy-conserved photon pairs. Furthermore, the multi-channel design leads to overlapped frequency regions for both signal and idler photons as shown in Fig. \ref{Fig5}(b), which cannot be implemented using orthogonal coding scheme with Hadamard codes, for example. This results from an unwanted cancellation between $c_1$ and $c_2$ in the common multiplexed frequency domain. 

Moreover, the multi-channel scheme can be more stable and resilient to channel errors, which we present in Figs. \ref{Fig5}(c) and \ref{Fig5}(d). As a demonstration, we design a corresponding single-channel setup in Fig. \ref{Fig5}(c) with $R$ segments and the same total code space of $M^R$ as in the multi-channel setup of Fig. \ref{Fig5}(a). To quantify the effect of $r'$ channel or segment errors, we use the length-$M$ nonorthogonal codes on all $R$ channels or segments separately using Eq. (\ref{C14}) and compare three cases of mismatched codes between the columns $(1,2)$ and $h=2$, the columns $(1,4)$ and $h=2$, and the columns $(1,4)$ and $h=4$. In this nonorthogonal code with quasi-orthogonality, the columns $(1,2)$ are orthogonal to each other, while the columns $(1,4)$ are not. We then calculate $g^{(2)}(0)$ for these three cases as shown in Fig. \ref{Fig5}(d), and they are normalized to the $g_{max}^{(2)}$ in the error-free case at $r'=0$ respectively for single-channel and multi-channel setups. We find a quadratic decrease in the single-channel setup as the number of mismatched channels increases, in contrast to a linear decrease in the multi-channel setup. This slow reduction of $g^{(2)}(0)$ manifests the tolerance of channel errors in the multi-channel nonorthogonal coding scheme which is potentially applicable in a reliable multiuser quantum communication.

There are two central advantages of the nonorthogonal coding scheme compared to the orthogonal one, which are (a) economical frequency ranges for coding and encoding protocols and (b) resilience of the multi-channel protocols under multiple channel errors. Our proposed nonorthogonal coding scheme promises a high-capacity operation and state engineering, wherein the enlarged Hilbert space can be used in a scalable quantum computation. For the facilitation of quantum cryptography protocols, the nonorthogonal coding scheme will be useful in multiuser quantum communication, where a reliable transmission can be achieved among many receivers under an influence of code corruptions as well as channel errors. Our proposed nonorthogonal coding scheme goes beyond and generalize the orthogonal one, which provides flexible and advantageous coding protocols at the price of less contrasted photon-photon correlations. 

\section{Conclusion and discussion}

Our results demonstrate the realization of encoding and decoding nonorthogonal codes on the spectrally-entangled photons in a multiplexing scheme. The photon pairs can be generated from the atomic cascade emissions with an advantageous telecom bandwidth which is best for an application of long-distance quantum communication. We specifically utilize a quasi-orthogonal space–time block code, which is a well developed technique in wireless classical communication and is useful for transmitting multiple copies of data sequences across many users. This nonorthogonal coding scheme can be applied in a multiplexing platform of entangled biphotons, where we show the performance of the codewords in terms of the second-order correlations. We further propose a multi-channel design to encode the nonorthogonal codes, where a high capacity of coding space is feasible. 

We have shown that the staircase structure of multiplexed photon pairs in a frequency-separated regime is the best dense structure that has the highest code space capacity with a limited frequency space. We also demonstrate the advantages of multi-channel nonorthogonal coding scheme which is more robust to channel errors than the single-channel setup using conventional orthogonal coding. This design is particularly applicable in quantum information processing and transmission, and is ready for multi-user quantum key distribution \cite{Gisin2002}. Moreover, our schemes can potentially integrate with a reconfigurable bandwidth allocation among multiple users \cite{Lingaraju2021}, which can further establish a practical scalable quantum networks, essential for an application of large-scale quantum computation. 

Finally, our proposed scheme can be readily implemented in squeezed quantum microcombs generated from the existing platform of silica microresonator on a silicon chip \cite{Yang2021}. Through the Kerr nonlinearity in the microresonator, the system generates broadband and energy-correlated photon pairs of signal and idler modes. With additional multiplexing as in our scheme for nonorthogonal coding, a multi-channel setup can be realized in these photon pairs. This platform can leverage the deterministic photon source with integrated photonics to enable compact and scalable quantum sources \cite{Mahmudlu2023}, and to offer new opportunities in continuous-variable quantum information processing \cite{Yang2021}.

\section*{ACKNOWLEDGMENTS}
We acknowledge support from the Ministry of Science and Technology (MOST), Taiwan, under the Grant No. MOST-109-2112-M-001-035-MY3 and No. MOST-111-2119-M-001-002. We are also grateful for support from TG 1.2 and TG 3.2 of NCTS at NTU. 
\appendix*
\section{Biphoton state}

Here we present the theoretical background of the spectrally-entangled biphoton state utilized in the multiplexing scheme in Fig. \ref{fig1:multiplexCoding}. Two pump fields $\Omega_a$ and $\Omega_b$ are applied in the atomic ensemble, where a signal field $\hat a_{\k_s,\lambda_s}$ and an idler field $\hat a_{\k_i,\lambda_i}$ can be generated via spontaneous emissions. The Hamiltonian in interaction picture can be written as \cite{Jen2016, Jen2017}
\bea
    V_I=&&-\sum_{m=1,2} \Delta_m \sum_{\mu=1}^{N} |m\rangle_{\mu} \langle m|-\sum_{m=a,b} \left(\frac{\Omega_m}{2}\hat{P}_m^{\dagger} + {\rm H.c.}\right)\nonumber\\
    &&-i\sum_{m=s,i}\left\{\sum_{\k_m,\lambda_m}g_m \hat{a}_{\k_m,\lambda_m} \hat{Q}_m^{\dagger}e^{-i\Delta \omega_m t}-{\rm H.c.}\right\},
\eea
where we set \(\hbar=1\) and define the detunings as \(\Delta_1 = \omega_a-\omega_1\), \(\Delta_2=\omega_a+\omega_b-\omega_2\). \(\omega_{a,b,s,i}\) are central frequencies of the pump and emitted photon fields with corresponding wave vectors \(\textbf{k}_{a,b,s,i}\). The collective dipole operators with a number of atoms $N$ are defined as \(\hat{P}^{\dagger}_a \equiv \sum_{\mu=1}^N |1\rangle_{\mu} \langle 0|e^{i \textbf{k}_a \cdot \textbf{r}_{\mu}}, \hat{P}^{\dagger}_b \equiv \sum_{\mu=1}^N |2\rangle_{\mu} \langle 1|e^{i \textbf{k}_b \cdot \textbf{r}_{\mu}}, \hat{Q}^{\dagger}_s \equiv \sum_{\mu=1}^N |2\rangle_{\mu} \langle 3|e^{i \textbf{k}_s \cdot \textbf{r}_{\mu}}, \hat{Q}^{\dagger}_i \equiv \sum_{\mu=1}^N |3\rangle_{\mu} \langle 0|e^{i \textbf{k}_i \cdot \textbf{r}_{\mu}}\), respectively. The coupling constant \(g_{m}\) has absorbed \((\epsilon_{\textbf{k}_m, \bar\lambda_m} \cdot \hat{d}^{*}_m)\), which involves the polarizations of quantized bosonic fields $\bar\lambda_m$ and the unit direction of the dipole operators $\hat d_m$.

To generate a spectrally-entangled photon pair \cite{Chaneliere2006}, the system is driven under weak excitations that \(\sqrt{N}|\Omega_a|\ll \Delta_1\). We then can express the state of the system using singly-excited Hilbert space \cite{Jen2016, Jen2017}, which reads
\bea
    |\psi(t)\rangle =&& \varepsilon(t)|0,vac\rangle+\sum_{\mu=1}^{N} A_{\mu}(t)|1_\mu,vac\rangle\nonumber\\
		&&+\sum_{\mu=1}^{N} B_{\mu}(t)|2_\mu,vac\rangle + \sum_{\mu=1}^{N} \sum_s C_s^{\mu}(t) |3_{\mu}, 1_{\textbf{k}_s,\lambda_s}\rangle \nonumber\\
		&&+ \sum_{s,i} D_{s,i}(t) |0,1_{\textbf{k}_s, \lambda_s}, 1_{\textbf{k}_i, \lambda_i} \rangle,
\eea
where single excitation states and vacuum states are \(|m_{\mu}\rangle=|m_{\mu}\rangle|0\rangle^{\otimes N-1}_{v\neq \mu}\) and \(|vac\rangle\), respectively. By applying Schrodinger equation \(i\hbar\frac{\partial}{\partial t}|\psi\rangle = V_I |\psi\rangle\), we obtain the equations of motion in a self-consistent form, 
\bea
i\dot{\varepsilon} =&& -\frac{\Omega_a^*}{2} \sum_{\mu=1}^N e^{-i\textbf{k}_a \cdot \textbf{r}_{\mu}} A_\mu ,\nonumber\\
i\dot{A_{\mu}} =&& -\frac{\Omega_a}{2} e^{i\textbf{k}_a \cdot \textbf{r}_{\mu}} \varepsilon - \frac{\Omega_b^*}{2}e^{-i\textbf{k}_b \cdot \textbf{r}_{\mu}}B_{\mu} - \Delta_1 A_{\mu}, \nonumber\\
i\dot{B_{\mu}} =&& -\frac{\Omega_b}{2} e^{i\textbf{k}_b \cdot \textbf{r}_{\mu}} A_{\mu}-\Delta_2 B_{\mu} - i \sum_{\textbf{k}_s, \lambda_s} e^{i\textbf{k}_s \cdot \textbf{r}_{\mu}} e^{-i\Delta \omega_s t} C_s^{\mu}, \nonumber\\
\dot{C}^{\mu}_s =&& i g_s^{*} e^{-i\textbf{k}_s \cdot \textbf{r}_{\mu}} e^{i\Delta \omega_s t} B_{\mu} - i \sum_{\textbf{k}_i, \lambda_i} g_i e^{-i\textbf{k}_i \cdot \textbf{r}_{\mu}} e^{i\Delta \omega_i t} D_{s,i}, \nonumber\\
i\dot{D_{s,i}} =&& ig_i^{*} \sum_{\mu=1}^N e^{-i\textbf{k}_i \cdot \textbf{r}_{\mu}} e^{i\Delta \omega_i t} C_s^{\mu}\nonumber,
\eea
where a system generates a strongly-correlated biphoton state under the four-wave mixing process with a generation rate $\propto |D_{s,i}|^2$. The detunings for the signal and the idler photons are \(\Delta \omega_s \equiv \omega_s - \omega_2 +\omega_3 -\Delta_2\) and \(\Delta \omega_i \equiv \omega_i - \omega_3\), respectively. 

Under the large detuning condition, $|\Delta_{1,2}| \gg |\Omega_{a,b}|$, we can solve the equations of motion under the adiabatic approximation and obtain the steady-state solutions. Then we further obtain, 
\begin{equation}
\begin{split}
A_{\mu} &\approx -\frac{\Omega_a}{2\Delta_1} e^{i\textbf{k}_a \cdot \textbf{r}_{\mu}}, \\
B_{\mu} &\approx \frac{\Omega_a \Omega_b}{4\Delta_1 \Delta_2} e^{i(\textbf{k}_a + \textbf{k}_b)\cdot \textbf{r}_{\mu}}.
\end{split}
\end{equation}
$C_{s}^{\mu}$ can be solved by the probability amplitude $C_{s, \textbf{k}_i}$ in a momentum space of $\k_i$, where $C_{s, \textbf{k}_i} = \sum_{\mu} C^{\mu}_s e^{-i \textbf{k}_i \cdot \textbf{r}_{\mu}}$. This leads to 
\bea
C_{s, \textbf{k}_i} = &&g_s^{*} \sum_{\mu} e^{i\Delta \textbf{k}\cdot \textbf{r}_{\mu}} \int^t_{-\infty} dt' e^{i \Delta \omega_s t'} \nonumber\\
&&\times e^{[-(\Gamma^N_3/2) + i\delta \omega_i](t-t')} b(t'),
\eea
where $b(t)=\frac{\Omega_a(t)\Omega_b(t)}{4\Delta_1\Delta_2}$. $\Gamma^N_3 = (N \bar{\mu}+1) \Gamma_3$ represents the superradiant decay rate for $|3\rangle \rightarrow |0\rangle$ transition, where $\bar{\mu}$ depends on the geometrical shape of the atomic ensemble. The cooperative Lamb shift is $\delta \omega_i=\int^{\infty}_0 d\omega \frac{\Gamma}{2\pi}[P.V.(\omega-\omega_3)^{-1}]N\bar{\mu}(\textbf{k})$, where $P.V.$ denotes a principal value and $\Gamma = |d|^2\omega^3/(3\pi \hbar \epsilon c^3)$ with $d$ the dipole moment for the transition.

Finally the probability amplitude $D_{s,i}(t)$ can be solved as 
\bea
D_{s,i}(t) = &&g_i^{*} g_s^{*} \sum_{\mu} e^{i\Delta \textbf{k}\cdot \textbf{r}_{\mu}} \int^t_{-\infty} \int^{t'}_{-\infty} dt'' dt' e^{i\Delta \omega_i t'} e^{i\Delta \omega_s t''} \nonumber\\
&&\times b(t'') e^{[-(\Gamma^N_3/2)+i\delta \omega_i](t'-t'')}.
\eea
When we choose the input pulses in Gaussian forms, $\Omega_a(t)=\frac{1}{\sqrt{\pi} \tau} \tilde{\Omega}_a e^{-t^2/\tau^2}, \Omega_b(t)=\frac{1}{\sqrt{\pi} \tau} \tilde{\Omega}_b e^{-t^2/\tau^2}$, with the pulse areas of Gaussian wave packets defined as \(\Tilde{\Omega}_a, \Tilde{\Omega}_b\), and their common pulse duration \(\tau\), the long-time limit integral as $t\rightarrow\infty$ gives 
\bea
D_{s,i} = \frac{\Tilde{\Omega}_a \Tilde{\Omega}_b g_i^{*} g_s^{*}}{4\Delta_1 \Delta_2} \frac{\sum^{N}_{\mu= 1}e^{i \Delta \textbf{k} \cdot \textbf{r}_{\mu}}}{\sqrt{2\pi}\tau} \frac{e^{-(\Delta \omega_s+\Delta \omega_i)\tau^2/8}}{\Gamma_3^N/2-i\Delta \omega_i},
\eea
where $N^{-1}\sum^{N}_{\mu= 1}e^{i \Delta \textbf{k} \cdot \textbf{r}_{\mu}}$ becomes one when $N\rightarrow\infty$, representing the four-wave mixing condition to generate highly-correlated biphoton state, that is \(\Delta\k\equiv\k_a+\k_b-\k_s-\k_i=0\). The steady-state wavefunction of the system can then be obtained as
\bea
    |\Psi\rangle \approx&& |0\rangle ^{\otimes N}  + \frac{\Tilde{\Omega}_a \Tilde{\Omega}_b g_i^{*} g_s^{*}}{4\Delta_1 \Delta_2} \frac{\sum^{N}_{\mu= 1}e^{i \Delta \textbf{k} \cdot \textbf{r}_{\mu}}}{\sqrt{2\pi}\tau} \nonumber\\
    &&\times \frac{e^{-(\Delta \omega_s+\Delta \omega_i)\tau^2/8}}{\Gamma_3^N/2-i\Delta \omega_i} |1_{\textbf{k}_s}, 1_{\textbf{k}_i} \rangle,\label{biphoton}
\eea
where the overall constant of the biphoton state in Eq. (\ref{biphoton}) represents the probability related to its generation rate. 


\end{document}